# Realization of topological superlattices and the associated interface states in one-dimensional plasmonic crystals

C. Liu and H.C. Ong[a]

Department of Physics, The Chinese University of Hong Kong, Shatin, Hong Kong, People's Republic of China

In analogous to the Su-Schrieffer-Heeger (SSH) model, one-dimensional (1D) electromagnetic (EM) crystals can exhibit nontrivial topological properties. In particular, when a nontrivial EM crystal is in contact with its trivial counterpart, a topologically protected interface state is formed. While much attention has been focused on single interface state, multiple interface states can interact collectively when under suitable conditions, giving rise to trivial or nontrivial band topologies resembling to the standard SSH model. Here, we study the topological properties of 1D metallic superlattices that support multiple interface states. We first demonstrate single interface state exists at the boundary between two topologically distinct metallic arrays that carry Bloch-like surface plasmon polaritons. Such interface state is then used as the building block for further constructing the superlattices. By exploiting the separation between two dimerized interface states and the distinct inter- and intra-interface configurations to facilitate different intracell and intercell interactions, we vary the band topology of the superlattices. More importantly, new superlattice interface state is formed when trivial and nontrivial superlattices are brought together. The superlattice interface state is found to have smaller angular divergence and longer localization length than its single counterpart, thus is desired for robust signal transmission and high finesse cavity.

[a] Email: hcong@phy.cuhk.edu.hk

## I. INTRODUCTION

Topological photonics has aroused much interest lately in realizing the so-called topological protected interface state [1-3]. When two photonic systems with distinct topological properties are brought together, an energy state must exist at the interface and spectrally lies within the band gap of the surrounding bulks [1-3]. Such bulk-edge correspondence yields the interface state which is robust against disorders and has become an attractive candidate for signal transmission [1-3]. In fact, various two-dimensional (2D) electromagnetic systems spanning from visible to microwave frequency have been demonstrated to support one-dimensional (1D) interface, or edge, state and backscattering free wave propagation has also been reported [3,4-10].

Other than 2D systems, 1D systems are also of interest because they support zero-dimensional interface state or soliton and serve as a toy model for studying the topological physics [11-16]. The Su-Schrieffer-Heeger (SSH) model, which originally was proposed to explain the charge transport properties of the conductive polyacetylene polymer, has provided the basic tool for studying topological insulators [18,19]. Its tight binding approach formulates the bipartite chains with different intracell and intercell interactions, leading to different Zak phases [20]. For the chain that possesses inversion symmetry, the Zak phase is quantized to 0 or $\pi$, defining whether it is topologically trivial or nontrivial [21]. For a nontrivial finite chain, two degenerate edge states, with energy lying within the band gap, are found at the terminations [20]. The model has successfully captured the importance of using Zak phase to characterize the band topology and the formation of edge states. Since then, SSH model has been generalized to include different elements such as long-range order [22,23], non-Hermiticity [24], higher dimensionality [16,25,26], Fano resonance [27], etc.

If one considers the interface state as a localized atomic orbital, it is possible to go beyond the standard SHH model by using two interface states as a building block for constructing superlattices. The superlattices may manifest different topological phases and, more importantly, new interface state. In contrast to the superlattices studied by Midya and Feng [28] and Wang et al [29] where multipartite basis is used to support higher order bands and multiple interface states, the superlattices proposed here involve bipartite interface states to form the basis. In fact, Munoz et al have theoretically studied such topological superlattices and found superlattice interface state (SIS) exists [30]. It is revealed SIS possesses longer localization length and is more resistance to on-site disorders than the regular interface state [30]. As a result, it may result in better wave propagation and less distortion in signal transmission.

Here, we demonstrate the SIS from 1D plasmonic superlattices. First, we find by varying the groove width of 1D Au plasmonic crystals (PmCs), band inversion is observed by angle-resolved reflectivity spectroscopy and verified by finite-difference time-domain (FDTD) simulation. In particular, the band topology can simply be determined by examining the spectral positions of the dark and bright coupled modes at the Brillouin zone center. When two topological trivial and nontrivial PmCs are bound together, an interface state is found within the bulk band gap. The interface state is then used as the building block for constructing the superlattices. By using a pair of interface states to form bipartite basis, we observe, analogous to the PmC counterparts, two dark and bright superlattice modes and band inversion can occur when varying the separation between two interfaces. We study the intracell and intercell coupling constants by temporal coupled mode theory (CMT), and find the constants are consistent with the SSH model in which stronger intercell coupling than that of intracell yields nontrivial properties. Two trivial and nontrivial superlattices are connected to demonstrate SIS. The SIS is shown to have longer localization length of 15.7 μm compared to 9.96 μm determined from the single interface state.

## II. BAND INVERSION AND ZAK PHASE OF PLASMONIC CRYSTALS

We first study the topological properties of 1D Au PmCs. The PmCs are fabricated by using focused ion beam (FIB) and their scanning electron microscopy (SEM) images are shown in the insets of Fig. 1(a) – (e), showing they are periodic rectangular grooves with period P = 600 nm, groove height H = 30 nm, and groove width W varying from 100 to 500 nm. After the sample preparation, they are then characterized by a homebuilt Fourier space optical microscope for measuring the angle- and wavelength-resolved reflectivity mapping [31,32]. Briefly, a supercontinuum white light laser is illuminated on the sample at a well-defined incident angle θ via the microscope objective lens and the reflection from the sample is collected by the same objective lens and a spectrometer-based CCD detector for spectroscopy [33]. By varying θ and at the same time measuring the reflectivity spectrum, we contour plot the reflectivity mapping. The p-polarized reflectivity mappings of the PmCs taken along the Γ-X direction are displayed in Fig. 1(a) – (e), showing the n = ±1 Bloch-like surface plasmon polaritons (SPPs), as indicated by two dash lines that are calculated by the phase matching equation $\frac{\varepsilon_{Au}}{\varepsilon_{Au}+1}\left(\frac{1}{\lambda}\right)^2 = \left(\frac{\sin\theta}{\lambda}+\frac{n}{P}\right)^2$, where $\varepsilon_{Au}$ is the dielectric constant of Au, are excited and they cross at normal incidence [34]. At the cross-point, the SPPs couple to form two dark and bright modes separated by an energy band gap [34-36]. The dark mode is nonradiative whereas the bright mode is radiative, giving rise to high and low reflectivity [36]. We plot the spectral positions of the dark and bright modes in Fig. 1(f) as a function of W. When W increases, one sees the band gap shrinks gradually, closes at W = 300 nm, and then reopens again afterwards. At the same time, the dark and bright modes switch places after the closure of the gap. We note the dark mode is a symmetry protected bound states in the continuum (BIC), featuring zero radiation damping [37]. As the dark and bright modes are associated with different SPP field symmetries, the interchange of two modes implies the occurrence of band inversion and the presence of a Dirac point at W = 300 nm [13,33,38]. More importantly, such band inversion also signifies the Zak phase of the -1 SPP band changes from 0 to $\pi$ or vice versa [12].

We perform FDTD simulations to further study the band topology. The p-polarized reflectivity mappings based on the same structures are provided in the Supplementary Information, and they agree with the experiment [33]. We then use the dipole source excitation method to calculate the electric near-field patterns of the -1 SPP band across the Brillouin zone from k = -$\pi$/P to $\pi$/P µm$^{-1}$ in the Supplementary Information [33]. The field patterns at the k = 0 and $\pi$/P µm$^{-1}$ for W = 100, 200, 400, and 500 nm PmCs are summarized in Fig. 2(a) – (d). At k = 0 µm$^{-1}$, we see the field symmetry in fact changes from odd to even with respect to the groove center when W increases, corresponding to the mode changes from dark to bright. On the other hand, at k = $\pi$/P µm$^{-1}$, the fields are even for all PmCs. Therefore, the field symmetries at two k points change for W = 100 and 200 nm PmCs but remain the same for W = 400 and 500 nm PmCs, yielding Zak phase = $\pi$ and 0 accordingly [37]. To directly determine the Zak phase, we employ the Wilson loop approach to calculate the Zak phase of the -1 SPP bulk band [39]. The Zak phase is defined as [21]:

$$\gamma_n = \int_{-\frac{\pi}{P}}^{\frac{\pi}{P}} X_{nn}(k) dk ,\qquad(1)$$

where $X_{nn}(k)$ is the Berry connection and given as [13,39]:

$$X_{nn}(k) = \frac{i \int_{unit\ cell} u_{nk}^*(x)\varepsilon(x)\frac{\partial u_{n,k}(x)}{\partial k}dx}{\int_{unit\ cell} u_{nk}^*(x)\varepsilon(x)u_{n,k}(x)dx},\qquad(2)$$

where $u_{n,k}(x)$ is the periodic function of the electric field components of the SPP Bloch wavefunctions and ε is the permittivity. The evolutions of individal phase difference, which is $X_{nn}(k)\Delta k$, of the -1 SPP band across the Brillouin zone with Δk = 0.04π/P are plotted in Fig 3(a) – (d), and the integrations give the Zak phases to be π for small W and 0 for larger W, in consistent with the results deduced from the field symmetry. We also have studied the Zak phases of the n = 0 SPP band, which is beyond the light line, by using the field symmetry approach for all PmCs and they are found to be π [33]. As the topology of the PmCs is defined by the sum of the Zak phases of all the isolated bands below the gap [12], we conclude the W = 100 and 200 nm PmCs are topological trival whereas W = 400 and 500 nm PmCs are nontrival.

## III.  SINGLE INTERFACE STATE

Once the topological properties of the PmCs have been studied, we then construct a PmC heterostructure with different topological phases on two sides to produce an interface state. We fabricate the heterostructure by FIB and the SEM image is displayed in the inset of Fig. 4(a), showing it has P and H = 600 and 40 nm and W = 100 (nontrivial) and 500 (trivial) nm on the left- and right-hand sides of the interface, which is indicated by the dash line. The p-polarized reflectivity mapping taken along the Γ-X direction is shown in Fig. 4(a), showing the +1 and -1 SPPs cross at normal incidence to yield two upper and lower bulk modes located at 610 and 640 nm and a band gap. However, unlike the regular PmCs, as two sides from the heterostructure carry different topological phases, a localized resonance with low reflectivity is clearly observed at λ ~ 622 nm within the band gap. The resonance has a narrow angular divergence of 2.1º if estimated from the FWHM of the angular profile [33]. For verification, we have simulated the reflectivity mapping of a similar finite heterostructure in Fig. 4(b), and it agrees with experiment that a localized mode is present at λ = 635 nm in the middle of the band gap together with two upper and lower bulk modes at 615 and 660 nm. We notice the mapping taken at positive and negative incident angles are slightly different due to the finite simulation size effect in which the numbers of unit cells on each side are limited to 40 to save the computational time. By examining the angular profile of the localized mode, the divergence is found to be 4.65º [33]. To check the localized mode is in fact the interface state, we simulate the near-field patterns in logarithmic scale at 20 nm above the heterostructure across the whole spectrum in Fig. 4(c) under normal incidence. The topological trivial and nontrivial PmCs are the left- and right-hand sides of the interface which is located at x = 0 μm. The fields are visible at the wavelengths corresponding to the upper and lower bulk modes as well as the localized mode. The fields observed from the bright mode, in which its wavelengths are 615 and 660 nm for the nontrivial and trivial PmCs, are always stronger than those from the dark counterparts because of the strong radiative coupling from the incidence to the resonance despite a lower quality factor [40]. More importantly, we find the localized mode is spatially localized at the interface and its fields decay rapidly into the bulk regions. The cross-section field profile at 635 nm is plotted in Fig. 4(d), showing and the fields decay asymmetrically and exponentially from the interface into two sides. The exponential decays follow well with the edge state from the SSH model [20,30] and their decay lengths are determined to be 4.17 and 3.45 μm for the trivial and nontrivial parts, respectively. We then Fourier transform the field pattern to k space in Fig. 4(d) and the localization length is deduced to be 9.96 μm.

**IV.    TWO COUPLED INTERFACE STATES**

The interface state can serve as the building block for assembling plasmonic superlattice. In fact, if each interface state can be treated as a single orbital, collective interaction between its neighboring interface states is then possible for producing continuous energy bands. Such interaction should be examined in prior to studying the superlattice. Therefore, we first demonstrate the interaction between two interface states by simulation [41,42]. We simulate two series of finite nT/mN/nT and nN/mT/nN tri-structures, where T and N denote trivial (W = 100 nm) and nontrivial (500 nm) PmCs, and n/m = 10/2, 9/4 and 8/6 are the numbers of PmC unit cells. All of them have a pair of interface states with different separations. In addition, two series have either N or T PmC as the middle layer. The corresponding p-polarized reflectivity mappings taken along the direction perpendicular to the grooves are illustrated in Fig. 5(a) and (b), showing, for all structures, two discrete interface coupled modes separated by an avoided crossing are seen within the bulk band gap. However, their characteristics are strongly system dependent. While they all show the size of the avoided crossing shrinks when the separation between two interfaces increases, the spectral positions of the bright and dark modes depend on the middle layer. Bright mode is observed at short wavelength in T/N/T but is flipped to long wavelength when N/T/N is used [42]. The near-field patterns of the bright and dark modes for 10T/2N/10T and 10N/2T/10N taken under normal incidence at 20 nm above the surface are calculated in Fig. 5(c) and (d). One sees the fields of the bright and dark modes are odd and even with respect to the center, indicating the field symmetries are strongly associated with the topological phase of the middle layer.

The interaction constant between two interface states can be determined by CMT in which the complex eigenfrequencies of the bright and dark modes are given as $\omega_\pm + i\Gamma_\pm = \omega_o + i\Gamma \pm (\kappa' + i\kappa'')$, where $\omega_o$ and $\Gamma$ are the angular frequency and decay rate of the uncoupled interface state and $\kappa'$ and $\kappa''$ are the real and imaginary parts of the interaction constant [33,43]. Given $\omega_+ - \omega_- = 2\kappa'$ and $\Gamma_+ - \Gamma_- = 2\kappa''$, we use the $\omega_\pm$ and $\Gamma_\pm$ of the bright and dark modes from Fig. 5(a) and (b) to determine $\kappa'$ and $\kappa''$ accordingly and the results are plotted in Fig. 5(e). While we see $\kappa'$ in both series decreases with increasing separation, it is negative in T/N/T but positive in N/T/N. Likewise, $\kappa''$ also exhibits opposite signs in two series despite their trend is similar. Therefore, the choice of the middle layer is likely to introduce a phase in the near- and far-field couplings between two interface states, governing both $\kappa'$ and $\kappa''$ [44]. We then calculate the moduli $|\kappa|$ of two series in Fig. 5(f), indicating they are almost the same and decrease with increasing the separation distance. As a result, we conclude the interface separation serves as a parameter for controlling different intracell and intercell interactions when formulating the bipartite unit cell in superlattice.

## V.    BIPARTITE SUPERLATTICES

We have fabricated two bipartite 2T/2N/2T and 1T/4N/1T superlattices by FIB and their SEM images are shown in the insets of Fig. 6(a) and (b). The T and N PmCs have 100 and 500 nm groove widths and identical P = 600 nm. Although two superlattices have the same period of 3.6 µm, their intra-interface separations are 1.2 and 2.4 µm. Therefore, following our earlier study on two coupled states, we expect they should exhibit different intercell and intracell interactions in which the intracell interaction is stronger than that of the intercell in 2T/2N/2T but is weaker in 1T/4N/1T. Their p-polarized reflectivity mappings taken along the Γ-X direction are shown in Fig. 6(a) and (b). In contrast to the heterostructure where only an isolated interface state is observed, two continuous dispersive bands arising from the collective interaction between the interface states are present within the bulk band gap [30]. Two interface bands cross and produce a new band gap in which its size is much smaller than that of the bulk counterpart [30]. In addition, two dark and bright modes are present, and their positions depend on the intra-interface separation. In analogy to the groove width dependence found in regular PmCs, we see the dark and bright modes are observed at short and long wavelengths when two interfaces in the unit cell are close to each other but switch places when they are farther apart. Band inversion may have occurred, signifying the superlattices carry different topological phases as well. The simulated reflectivity mappings of two superlattices are provided in the Supplementary Information for reference [33].

To examine the band topology of the lower interface band, we have simulated the near-field patterns of two superlattices across the first Brillouin zone in the Supplementary Information [33]. Their field patterns at the zone center and boundary are summarized in Fig. 6(c) and (d), showing the field symmetries are both even in 2T/2N/2T superlattice but are even and odd in 1T/4N/1T, facilitating 0 and $\pi$ Zak phases. To double check, we use the Wilson loop method to determine the Zak phases. The individual phase differences are plotted in Fig. 6(e) and (f) as a function of k and the integrals yields Zak phase = 0 and $\pi$ for 2T/2N/2T and 1T/4N/1T superlattices, respectively, and they are consistent with the field symmetry results. We also see the interface bands can interact with the bulk bands to form avoided crossings at $\theta \sim \pm 4.4^\circ$. The crossing in 1T/4N/1T is larger than that in 2T/2N/2T, indicating different interactions between the interface and bulk bands.

## VI. TEMPORAL COUPLED MODE THEORY

As the occurrence of band inversion in two superlattices is determined by the interplay between intercell and intracell interactions, we attempt to estimate the coupling constants by using CMT [45-47]. In analogy to the SSH model, two interface states form a dimerized basis. The dynamics of two localized modes at site s, $u_s$ and $v_s$, can be approximated as:

$$\frac{du_s}{dt} = -i\tilde{\omega}_o u_s + i\kappa_1 v_s + i\kappa_2 v_{s-1}$$
$$\frac{dv_s}{dt} = -i\tilde{\omega}_o v_s + i\kappa_1 u_s + i\kappa_2 u_{s+1}, \qquad (3)$$

where $\tilde{\omega}_o$ is the complex resonant frequency, and $\kappa_1$ and $\kappa_2$ are the complex intracell and intercell coupling constants, respectively. Note that Eq. (3) is a tight binding approach and only the nearest neighbor interaction is considered. For a periodic system, we apply the boundary condition such that $u_{s+1} = u_s e^{ikP} e^{-i\omega t}$ and $v_{s-1} = v_s e^{-ikP} e^{-i\omega t}$, where k and $\omega$ are momentum and angular frequency. As a result, Eq. (3) becomes:

$$i \begin{bmatrix} \omega - \tilde{\omega}_o & \kappa_1 + \kappa_2 e^{-ikP} \\ \kappa_1 + \kappa_2 e^{ikP} & \omega - \tilde{\omega}_o \end{bmatrix} \begin{bmatrix} u_s \\ v_s \end{bmatrix} = 0. \qquad (4)$$

By solving the determinant of Eq. (4) to be zero, the complex eigenfrequency can be expressed as:

$$\tilde{\omega}_\pm = \omega_\pm + i\Gamma_\pm = \tilde{\omega}_o \pm \sqrt{\kappa_1^2 + \kappa_2^2 + 2\kappa_1\kappa_2 \cos(kP)}. \tag{5}$$

In fact, Eq. (5) can be used to fit the spectral positions and the linewidths of the upper and lower interface bands near normal incidence in Fig. 6(a) and (b) to determine $\kappa_1$ and $\kappa_2$.

The spectral positions and the linewidths of the upper and lower bands are plotted against k in Fig. 7(a) and (b) and they are best fitted with Eq. (5) as shown by the dash lines. It is noted that two bands are not isolated but interacted with the Rayleigh's anomalies, as indicated by the dash lines in Fig. 6(a) and (b), thus modifying the decay rate profiles and making them less smooth with k. Nevertheless, the $\kappa_1$ and $\kappa_2$ are roughly determined to be -0.075 - i0.0047 and 0.063 + i0.0046 for 2T/2N/2T and -0.063 – i0.0052 and 0.07 + i0.0032 for 1T/4N/1T. The corresponding moduli $|\kappa_1|$ and $|\kappa_2|$ are calculated to be 0.075 and 0.063 for 2T/2N/2T and 0.063 and 0.07 for 1T/4N/1T. In consistent with the SSH model [20], $\kappa_2$ is larger than $\kappa_1$ in 1T/4N/1T superlattice where π Zak phase is obtained but is smaller in 2N/2T/2N to yield 0 Zak phase. As a result, we conclude 1T/4N/1T superlattice is topological nontrivial whereas 2T/2N/2T is trivial.

## VII. SUPERLATTICE INTERFACE STATE

Once the superlattices that exhibit different band topologies are available, we connect them to realize SIS [30]. The inset of Fig. 8(a) shows the SEM image of the superlattice heterostructure which consists of 1T/4N/1T (nontrivial) and 2T/2N/2T (trivial) on the left- and right-hand sides of the interface, as indicated by the dash line. The p-polarized reflectivity mapping taken along the Γ-X direction is shown in Fig. 8(a). A localized mode with lower reflectivity than its vicinities is observed at λ = 635 nm at normal incidence close of the superlattice bands. Because the size of the band gap is small, the mode is difficult to be distinguished from the bands. The mode has a much narrower angular divergence of 1.64° than that of the PmC interface state [33]. We have simulated the reflectivity mapping of the heterostructure is shown in Fig. 8(b), and it agrees with the experiment that a low reflectivity localized mode is present at 635 nm. The angular divergence is found to be 2.29° [33]. To verify the localized mode is in fact SIS, we also have simulated the near-field pattern mapping taken at 20 nm above the surface under normal incidence across the spectrum in Fig. 8(c), showing the fields are strong at three spectral regions which are 630, 635, and 648 nm. While the fields at 630 and 648 nm correspond to the bulk bands of the heterostructure, the field at 635 nm is strongly localized at the interface at x = 0 μm and it decays rapidly into the bulk regions. The cross-section field profile at 635 nm is then plotted in logarithmic scale in Fig. 8(d) to show the field decays exponentially and asymmetrically from the interface, and their decay lengths are estimated to be 26.7 and 26 nm in the trivial and nontrivial sides, respectively. We Fourier transform the field pattern in Fig. 8(d) and the localization length is determined to be 15.7 μm, which is longer than that of the single interface state.

## VIII. CONCLUSION

In summary, we have demonstrated topologically protected interface state and superlattice interface state (SLS) in 1D Au plasmonic crystals (PmCs) and superlattices. The topological properties of the PmCs have been studied by angle- and wavelength-resolved reflectivity spectroscopy and FDTD simulations. For PmCs, it is found the topological phase can be changed from trivial to nontrivial by varying the groove width. The spectral positions of the dark and bright coupled modes at the Brillouin zone center serve as a good indicator to reveal the field symmetry as well as the Zak phases of the plasmonic bulk bands. Well defined interface state is observed within the bulk band gap at the boundary between trivial and nontrivial PmCs both experimentally and numerically. Once the interface state is available, it is then used as the building block for constructing superlattices. We then show the interaction between two interface states strongly depends on their separation and thus the intracell and intercell interactions in bipartite superlattice can be engineered by varying the intra- and inter-interface distances. Topologically trivial and nontrivial superlattices have been fabricated and their characteristics are consistent with the SSH model. Finally, two superlattices carrying distinct topological phases are connected to demonstrate SIS. The SIS is shown to have longer localization length of 15.7 μm when compared to the 9.96 μm from the single interface state. As SLS is more resistance to on-site disorders than the regular interface state, our results may provide an approach for more robust signal transmission and higher finesse cavity.

## IX. ACKNOWLEDGEMENT

This research was supported by the Chinese University of Hong Kong through Area of Excellence (AoE/P-02/12) and Innovative Technology Funds (ITS/133/19 and UIM/397).

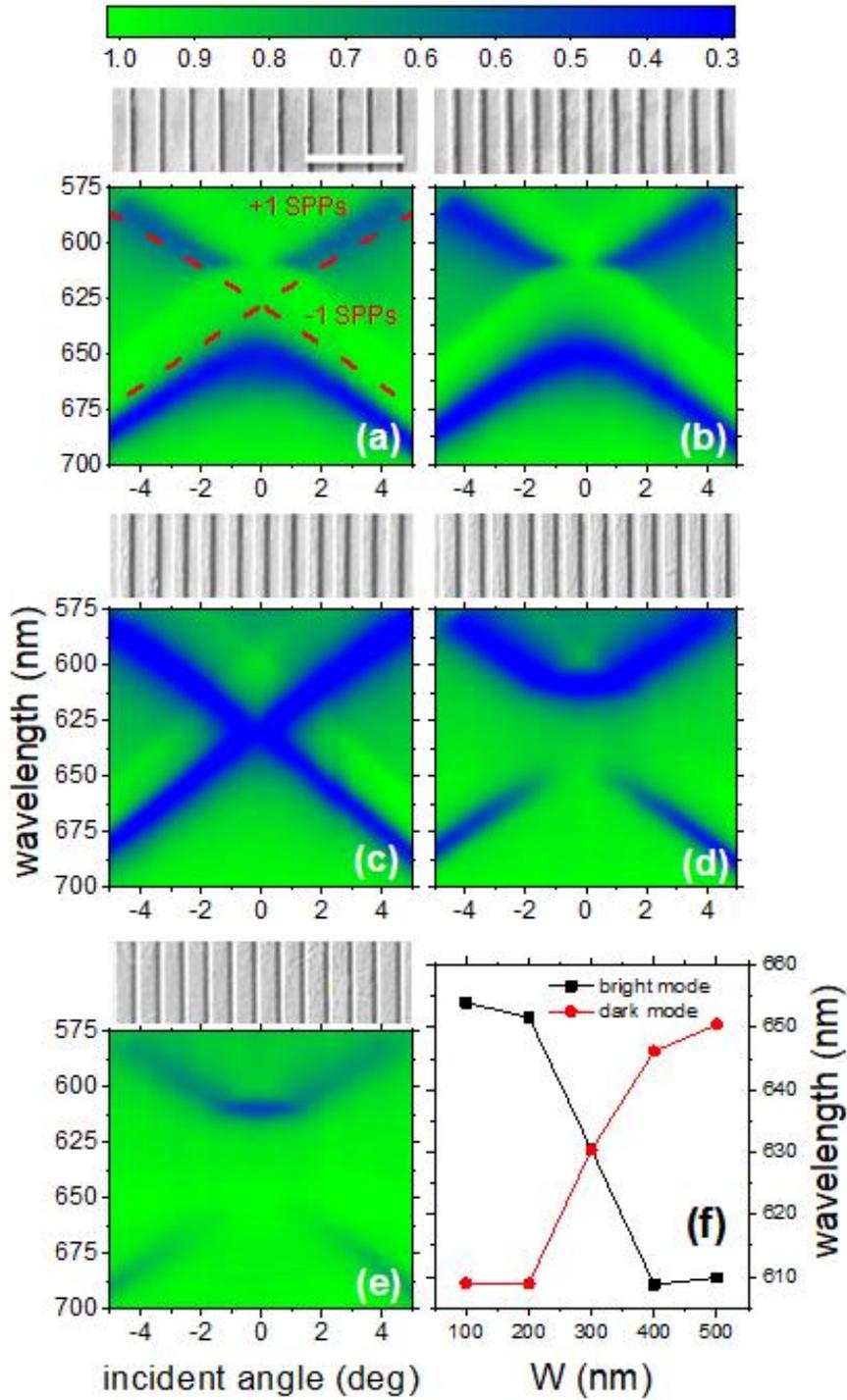

Fig. 1. (a) – (e) The angle- and wavelength-resolved reflectivity mappings of the PmCs with W = 100, 200, 300, 400, and 500 nm taken along the Γ-X direction. The +1 and -1 SPPs calculated by the phase matching equation are indicated by the dash lines. Inset: the SEM images of the PmCs and the scale bar is 2 μm. (f) The plot of the spectral positions of the dark and bright modes.

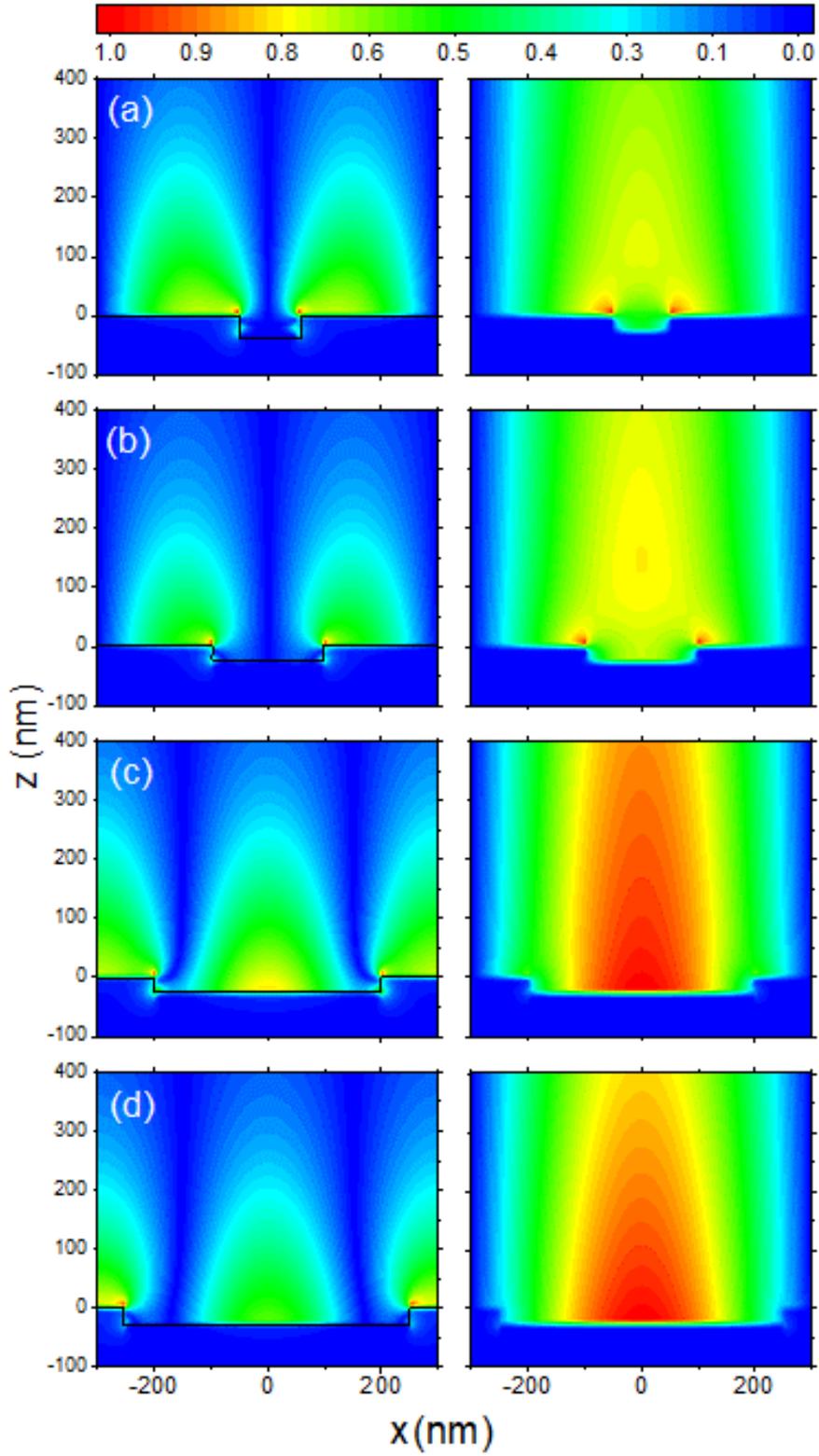

Fig. 2. (a) – (d) The electric near-field, the modulus of the surface normal component $|E_z|$, patterns of W = 100, 200, 400, and 500 nm PmCs calculated at the Brillouin zone center (left) and boundary (right) for the -1 SPP band using dipole source excitation method.

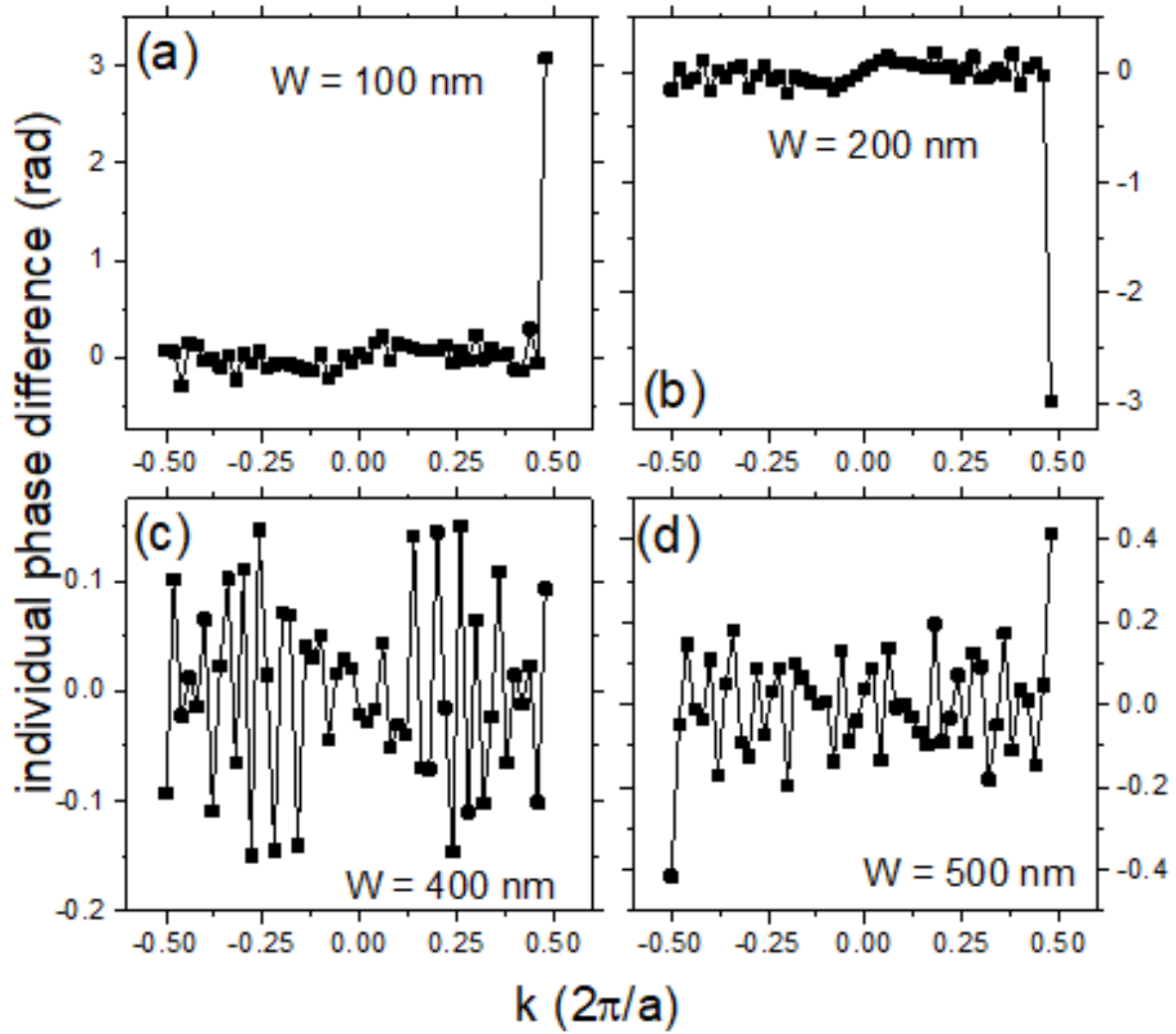

Fig. 3. (a) – (d) The plots of individual phase difference as a function of k for W = 100, 200, 300, and 400 nm PmCs determined by the Wilson loop method. The integrations yield the Zak phases = π for W = 100 and 200 nm PmCs and = 0 for W = 300 and 400 nm PmCs.

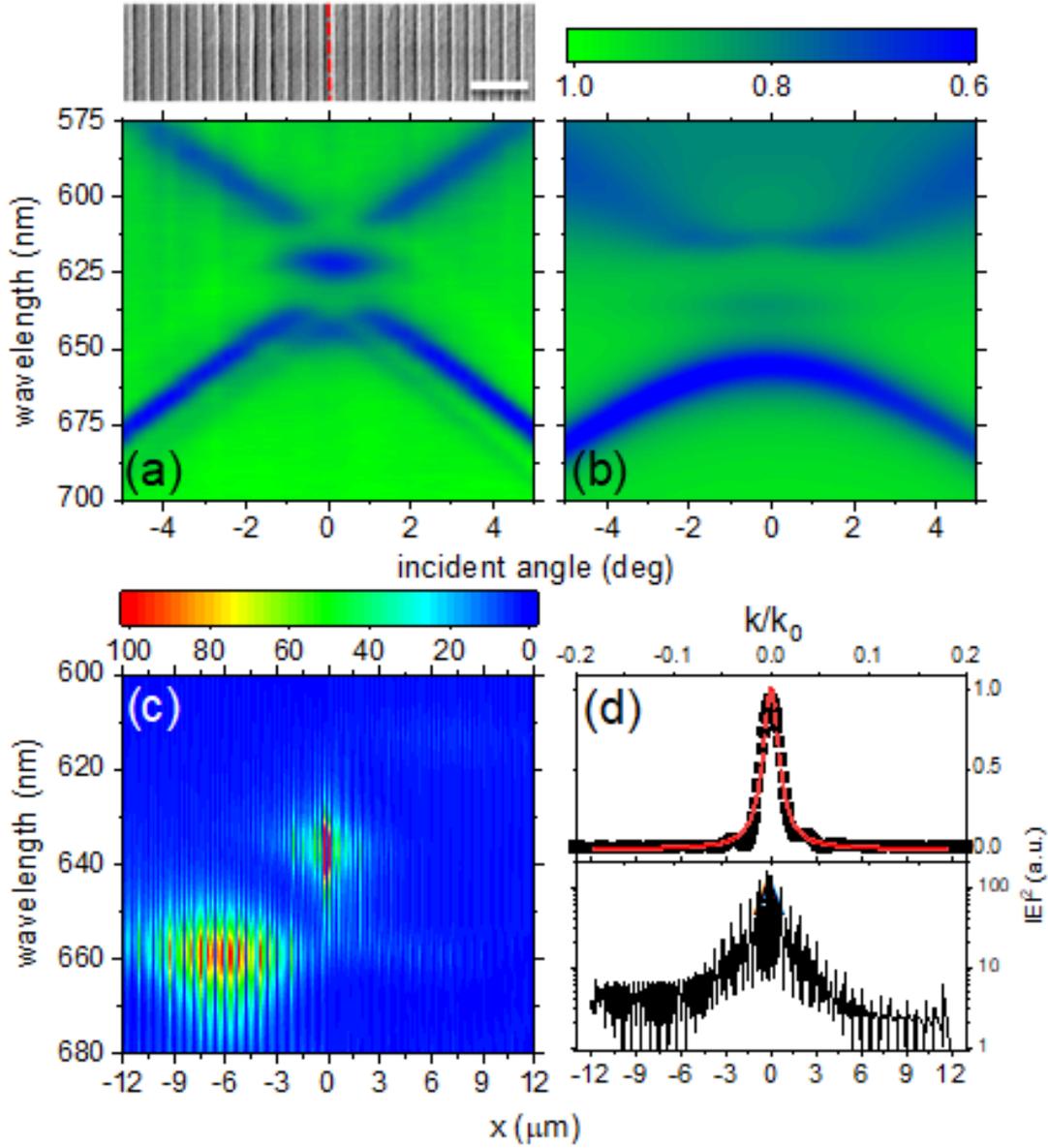

Fig. 4. The (a) experimental and (b) simulated reflectivity mappings of the heterostructure taken along the Γ-X direction, showing an interface state is found within the gap at λ = 635 nm. The inset shows the SEM image, and the interface is indicated by the dash line. The scale bar is 2 μm. (c) The near-field intensity, $|E|^2$, mapping in logarithmic scale taken at 20 nm above the heterostructure across the whole spectrum under normal incidence. The interface is located at x = 0 μm. (d) The near-field intensity profile of the interface state taken at λ = 635 nm and the corresponding Fourier transformed profile. The dash line is the Lorentzian fitting for determining the localization length.

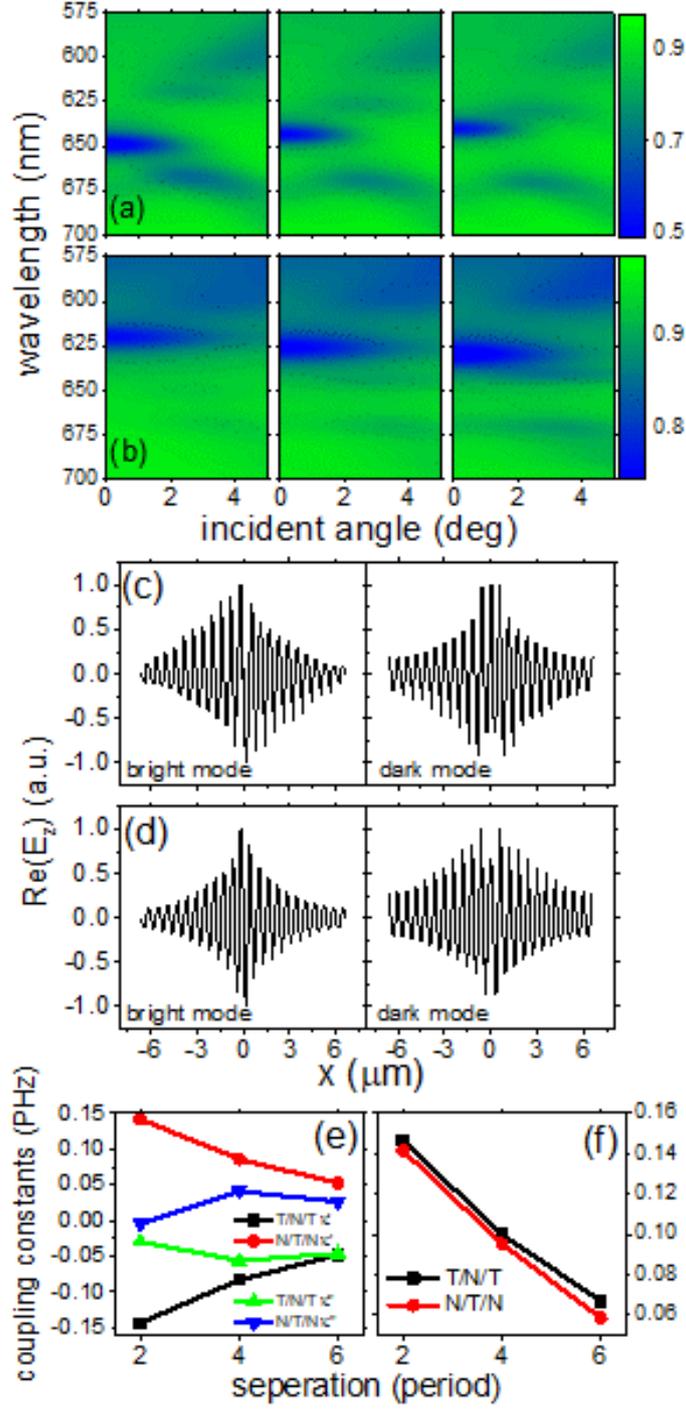

Fig. 5. The reflectivity mappings of two series of (a) nT/mN/nT and (b) nN/mT/nN tri-structures. For each series, n/m = 10/2, 9/4 and 8/6 for the left to the right. The real part of the surface normal near-field patterns, Re(E$_z$), of the bright and dark modes for (c) 10T/2N/10T and (d) 10N/2T/10N tri-structures. The (e) complex coupling constants $\kappa'$ and $\kappa''$ and their (f) moduli $|\kappa|$ are plotted as a function of separation between two interfaces.

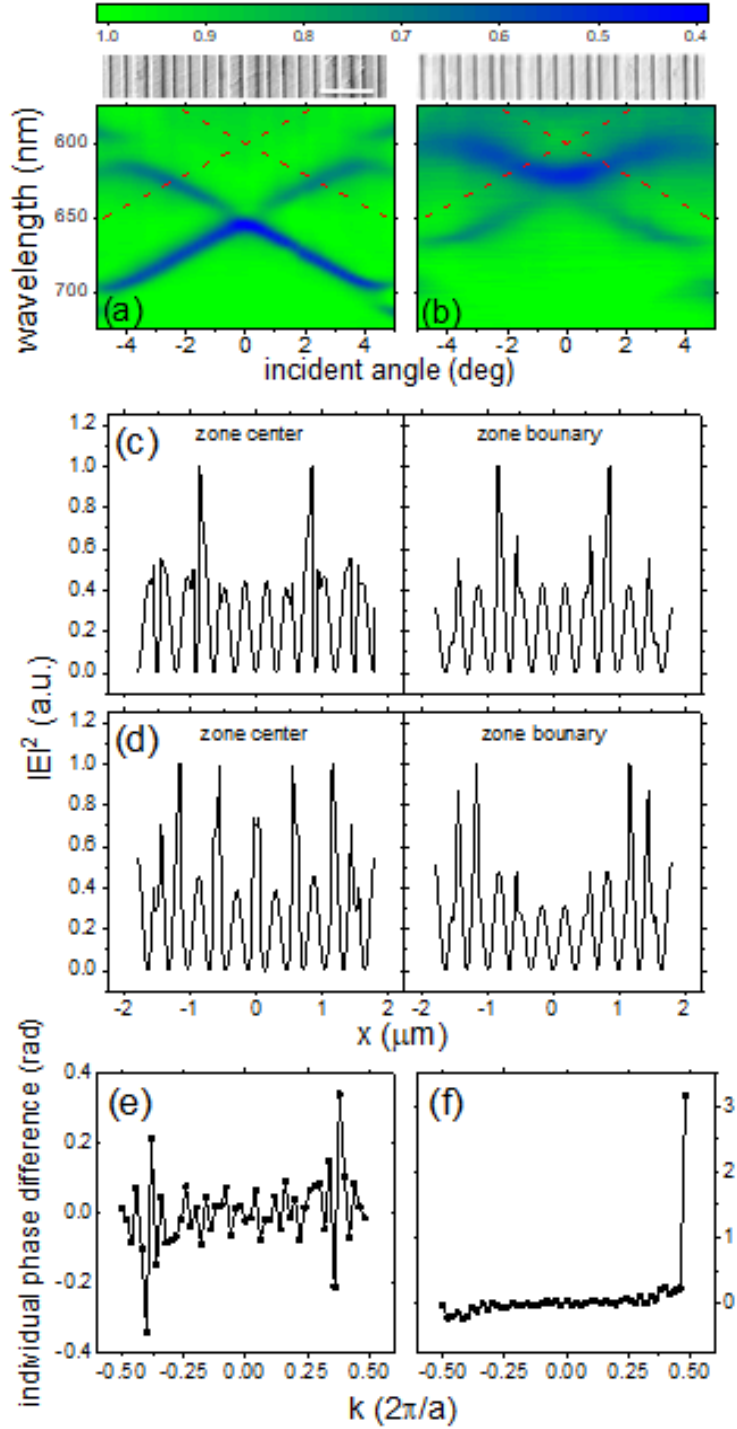

Fig. 6. The reflectivity mappings of two bipartite (a) 2T/2N/2T and (b) 1T/4N/1T superlattices taken along the Γ-X direction. The dash lines are the Rayleigh's anomalies. Inset: the SEM images of two superlattices and the scale bar is 2 μm. The near-field intensity, $|E|^2$, patterns of (c) 2T/2N/2T and (d) 1T/4N/1T superlattices taken at the zone center and boundary. The plots of individual phase difference as a function of k for (e) 2T/2N/2T and (f) 1T/4N/1T superlattices, yielding 0 and π Zak phases.

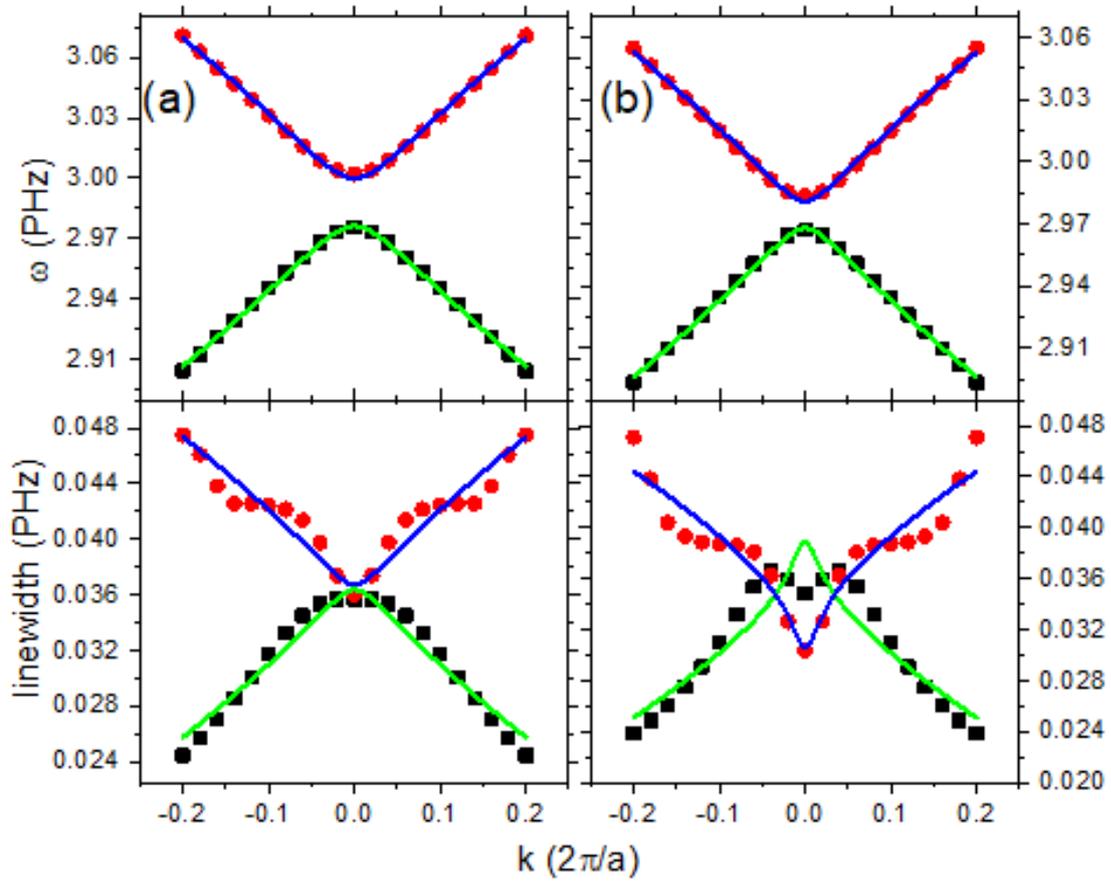

Fig. 7. The plots of the spectral positions and the linewidths of the upper and lower bands for (a) 2T/2N/2T and (b) 1T/4N/1T superlattices. The dash lines are the best fits by using temporal CMT.

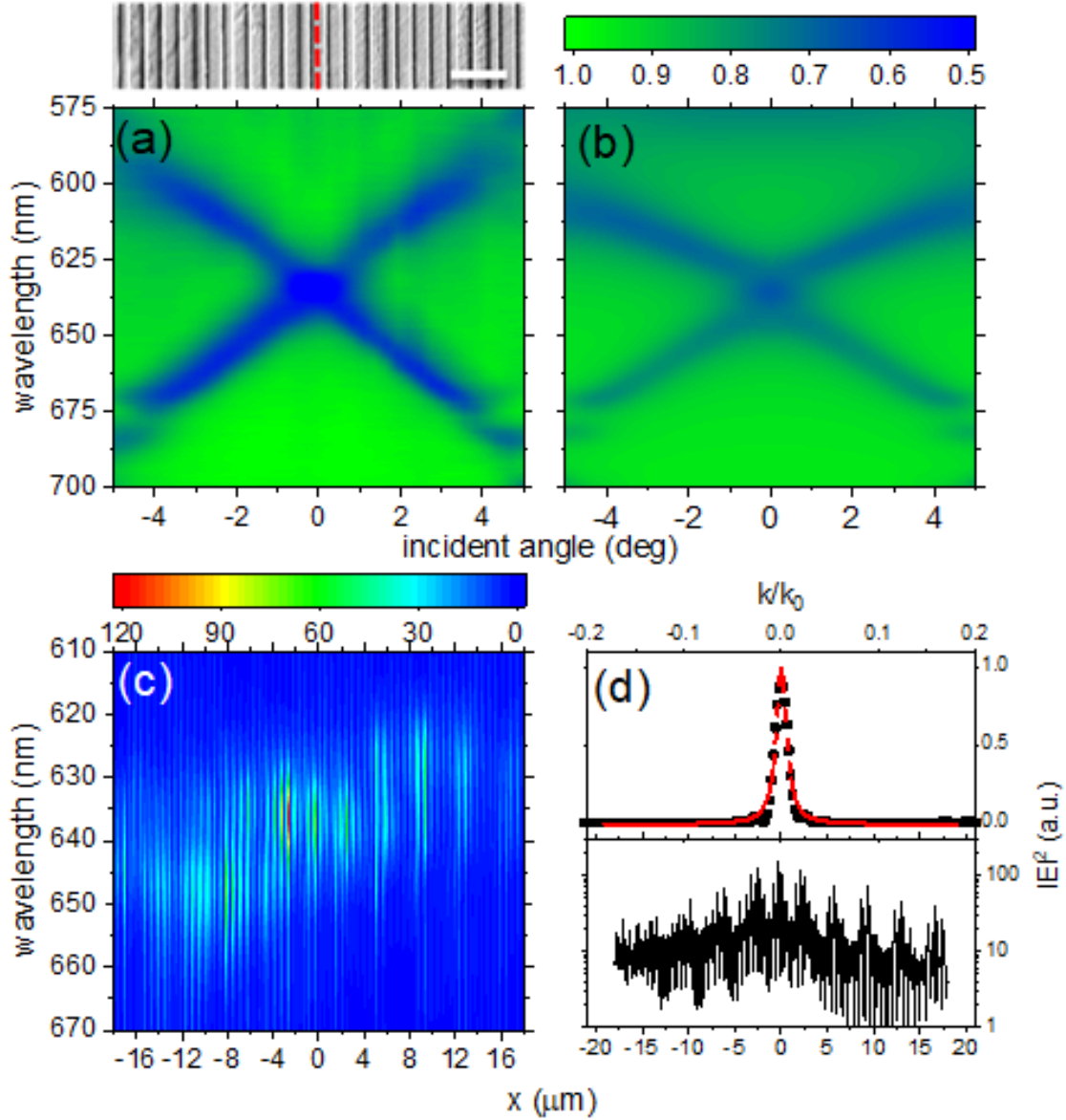

Fig. 8. (a) The reflectivity mapping of the superlattice heterostructure taken along the Γ-X direction, showing an interface state with low reflectivity is formed between two interface bands at λ = 635 nm. Inset: the SEM images of the superlattice heterostructure and dash line is the interface. The scale bar is 2 μm. (b) The simulated reflectivity mapping of the superlattice heterostructure. (c) The near-field mapping in logarithmic scale taken at 20 nm above the heterostructure across the whole spectrum under normal incidence. The interface is at x = 0 μm. (d) The near field intensity, $|E|^2$, profile of the superlattice interface state at λ = 635 nm and the corresponding Fourier transformed profile. The Lorentzian fit is shown as the dash line for determining the localization length.

# Supplementary Information

**Realization of topological superlattices and the associated interface states in one-dimensional plasmonic crystals**

C. Liu and H.C. Ong

Department of Physics, The Chinese University of Hong Kong, Shatin, Hong Kong, People's Republic of China

### A. Schematic of the Fourier space optical microscope for angle- and wavelength resolved reflectivity mapping

Fig. S1 shows the schematic of the Fourier space optical microscope. Briefly, a broadband supercontinuum laser from a photonic crystal fiber is collimated and then passed through a set of polarizers, wave plates, and lenses before being focused onto the back focal plane (BFP) of a 100X objective lens (OB) with numerical aperture = 0.9. By displacing the focused spot across the BFP of the objective lens using a motorized translation stage, the light exiting from the objective lens is collimated again and the incident polar angle θ onto the sample is given by sinθ = *d/f*, where *d* is the distance between the focused spot and the optical axis of the objective lens and *f* is the focal length of the objective. In addition, the azimuth angle ϕ can be varied by a motorized rotation sample stage to align the incident plane to the Γ-X direction of the array. The light reflected from the array is then collected by the same objective lens and passes through another set of analyzers and lenses before being detected by a spectrometer-based CCD detector.

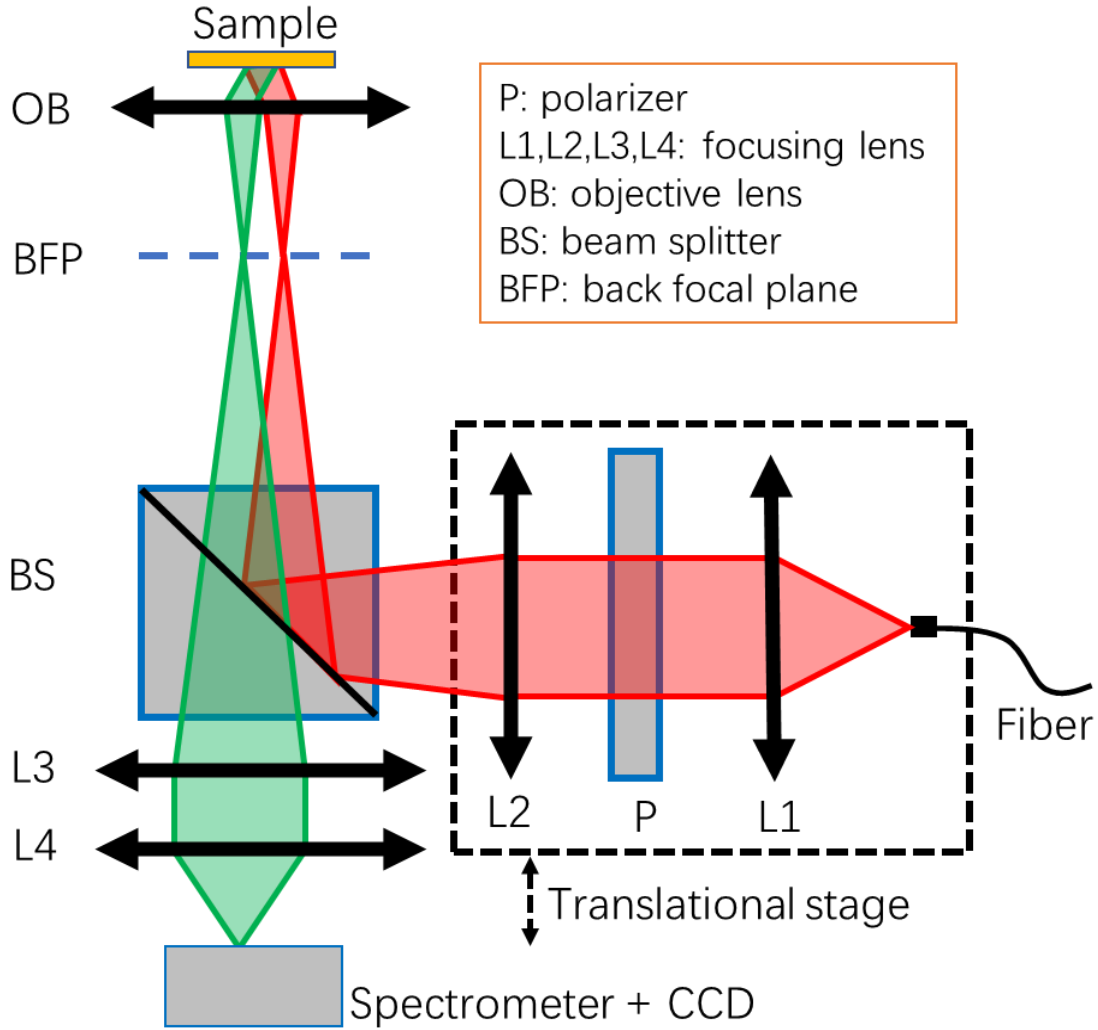

Fig. S1. The schematic of the Fourier optical microscope.

**B. Finite-difference time-domain simulation (FDTD) of reflectivity mappings of 1D Au plasmonic crystals (PmCs)**

FDTD has been used to simulate the p-polarized angle- and wavelength-resolved reflectivity mappings of different 1D Au PmCs along the Γ-X direction (Fig. S2(b) – (f)). The unit cell is illustrated in Fig. S2(a), and it has period P = 600 nm, groove height H = 30 nm and groove width W varying from 100 to 500 nm with a step size = 100 nm. The PmCs are immersed in air and the dielectric constant of Au is obtained from Ref. [1]. Bloch boundary condition is used at two sides and perfectly matched layer is used at the top and bottom.

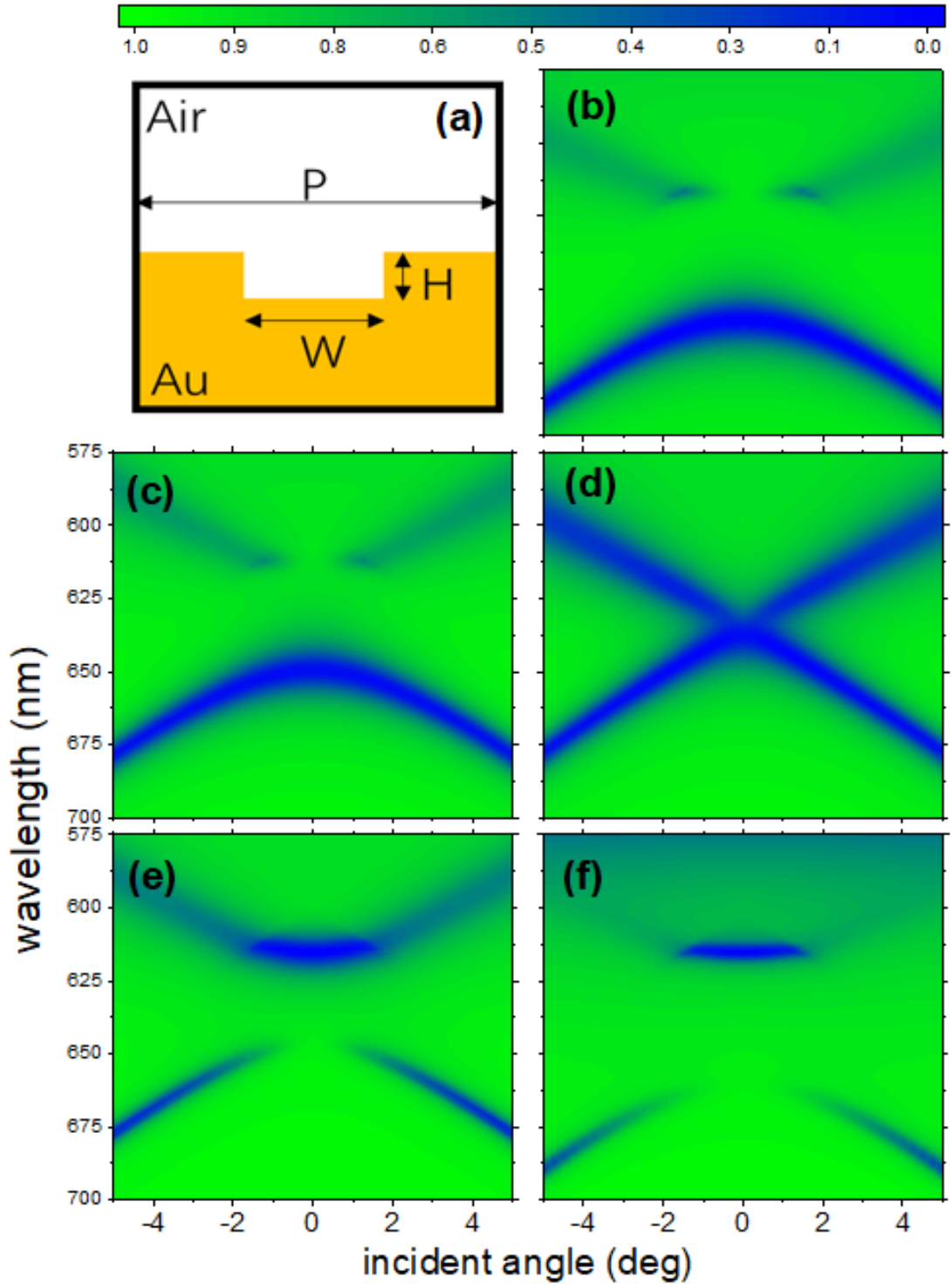

Fig. S2. (a) The FDTD simulation cell. The FDTD simulated reflectivity mappings for W = (b) 100, (c) 200, (d) 300, (e) 400, and (f) 500 nm PmCs.

**C. Simulated near-field patterns of the -1 surface plasmon polariton (SPP) band of 1D PmCs across the first Brillouin zone**

By using the dipole source excitation method, the complex near-field patterns of the -1 SPP band of 1D Au PmCs have been simulated. The real and imaginary parts of the surface normal components, Re($E_z$) and Im($E_z$), taken at 20 nm above the surface across the Brillouin zone from k = -π/P to π/P μm$^{-1}$ are shown in Fig. S3 for W = 100, 200, 400, and 500 nm PmCs. They will then be used for determining the Zak phase by the Wilson loop method.

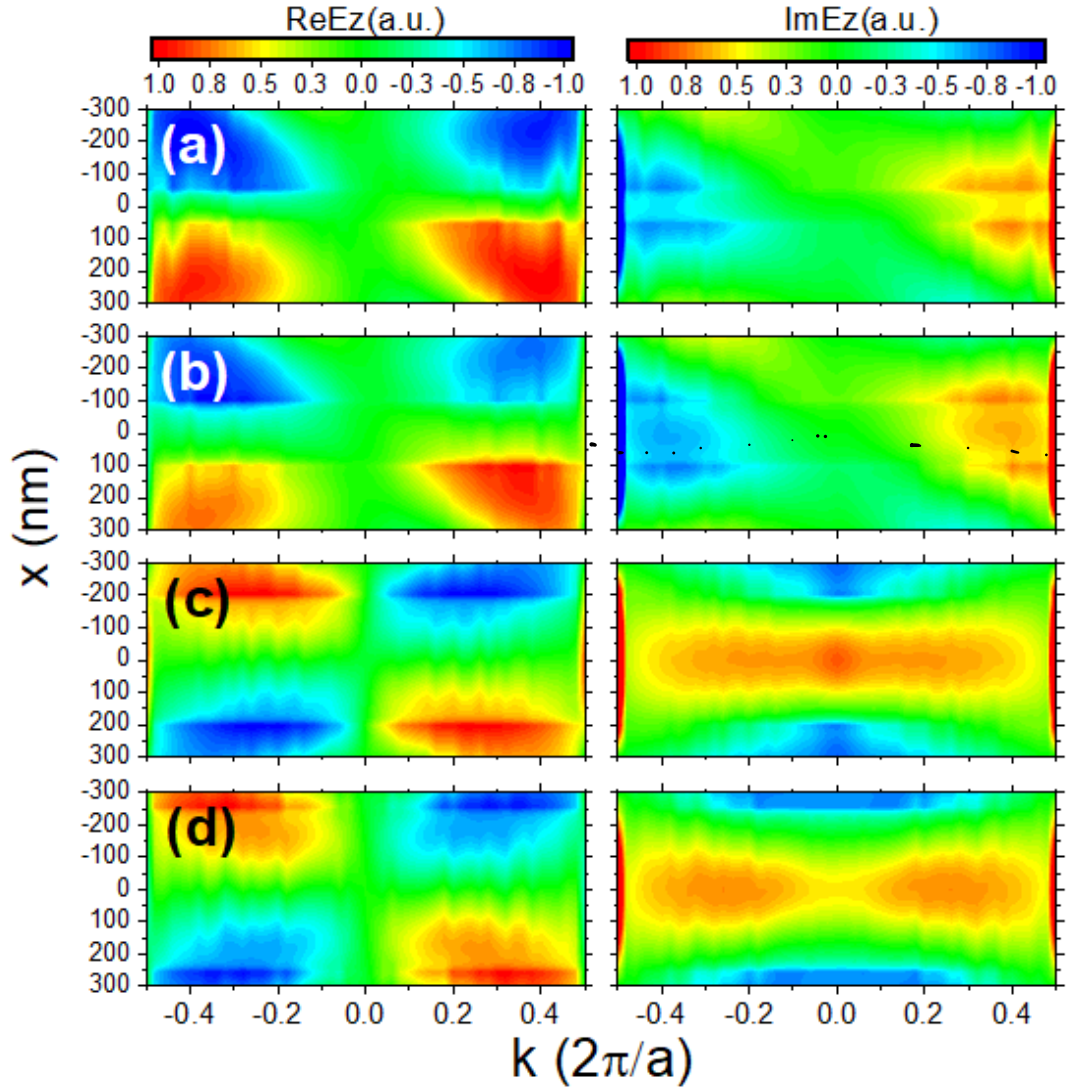

Fig. S3. The (left) real and (right) imaginary parts of the surface normal components, $E_z$, across the Brillouin zone from k = -π/P to π/P μm$^{-1}$ for W = (a) 100, (b) 200, (c) 400, and (d) 500 nm PmCs.

### D. Simulated near-field patterns of the 0 SPP band at the zone boundary

The near-field, the modulus of surface normal component, patterns of the 0 SPP bands of 1D PmCs for W = 100, 200, 400 and 500 nm taken at the zone boundary are shown in Fig. S4, showing the fields all have odd symmetry. On the other hand, the fields at the zone center, which has zero energy and thus longer wavelength, are argued to be even. As a result, the Zak phases for the 0 SPP bands are $\pi$ for all PmCs.

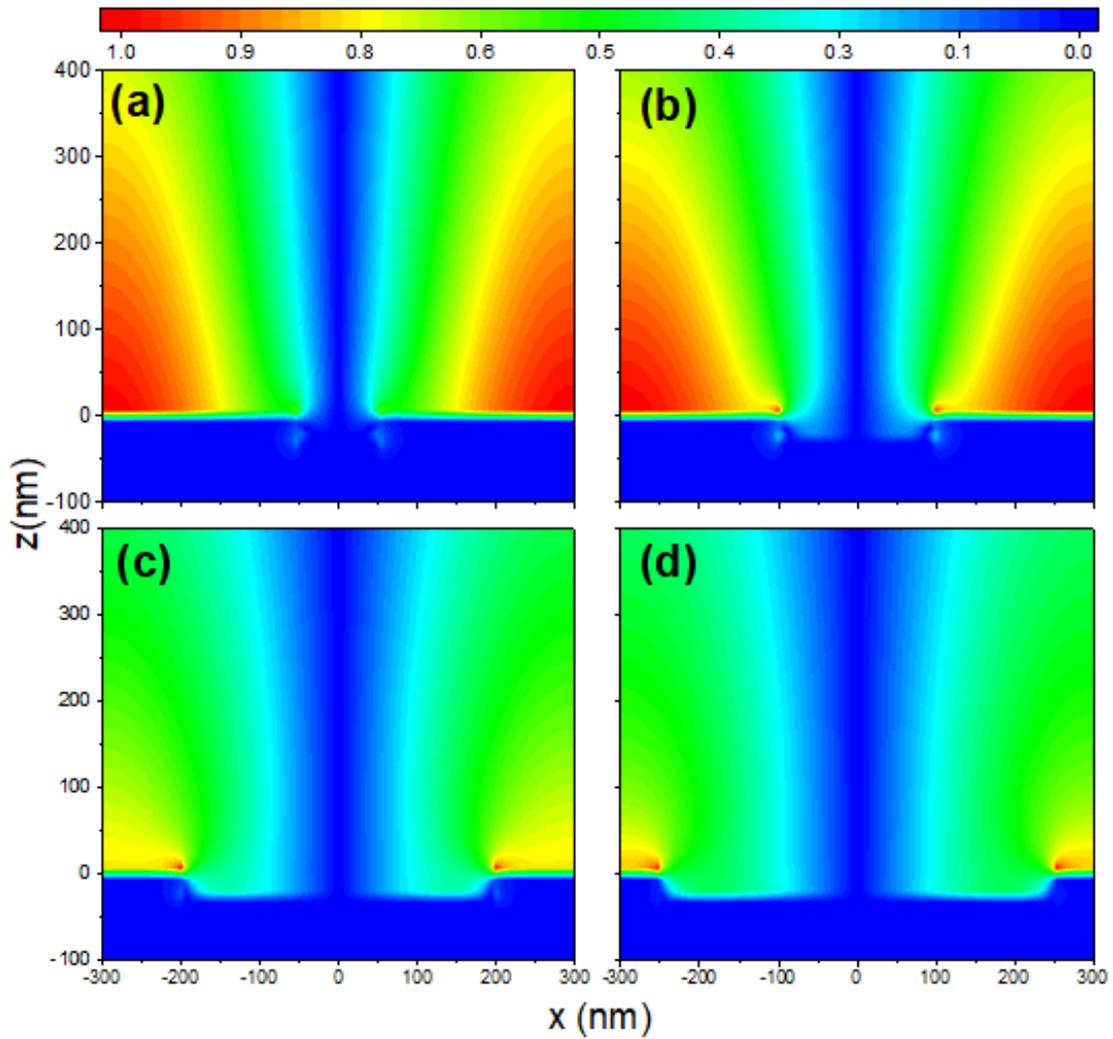

Fig. S4. The $|E_z|$ near-field patterns of the 0 SPP bands for W = (a) 100, (b) 200, (c) 400, and (d) 500 nm PmCs taken at the zone boundary.

### E. Angular profile of the single interface state

The experimental angular profile of the interface state in the 1D heterostructure taken at $\lambda$ = 622 nm is shown in Fig. S5(a). It is fitted with a Lorentzian function to determine the FWHM to be $\Delta\theta$ = 2.076 deg. The simulated angular profile of the interface state in the 1D heterostructure is shown in Fig. S5(b). It is fitted with a Lorentzian function to determine the FWHM to be $\Delta\theta$ = 4.65 deg.

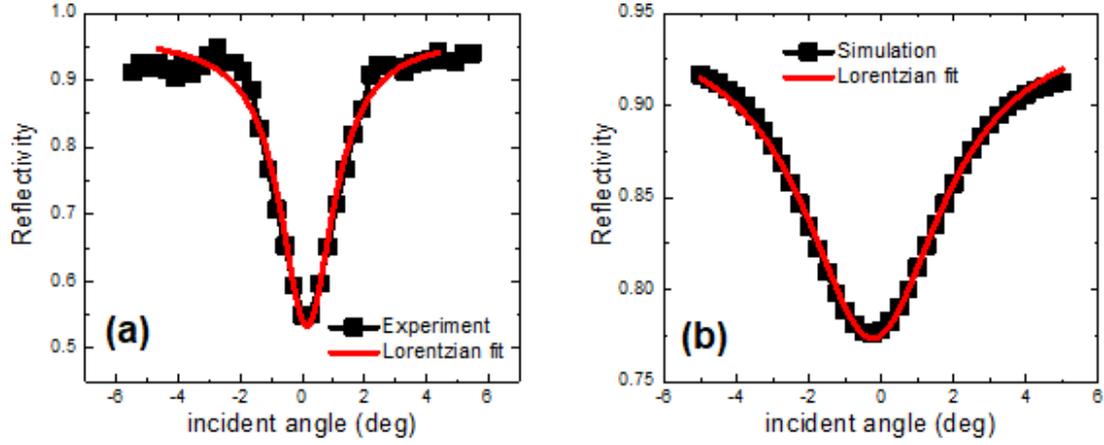

Fig. S5. The (a) experimental and (b) simulated angular profiles of the interface state.

## F. Simulated reflectivity mappings of 2T/2N/2T and 1T/4N/1T bipartite superlattices

The unit cells of 2T/2N/2T and 1T/4N/1T bipartite superlattices are shown in Fig. S6(a). Both contain six PmCs with P = 600 nm and H = 30 nm, making the period to be 3.6 μm. The T (trivial) and N (nontrivial) PmCs have W = 100 and 500 nm, respectively. They have a pair of interfaces and the intra-interface separations are 1.2 and 2.4 μm.

The corresponding p-polarized reflectivity mappings taken along the Γ-X direction of two superlattices are shown in Fig. S6(b) and (c).

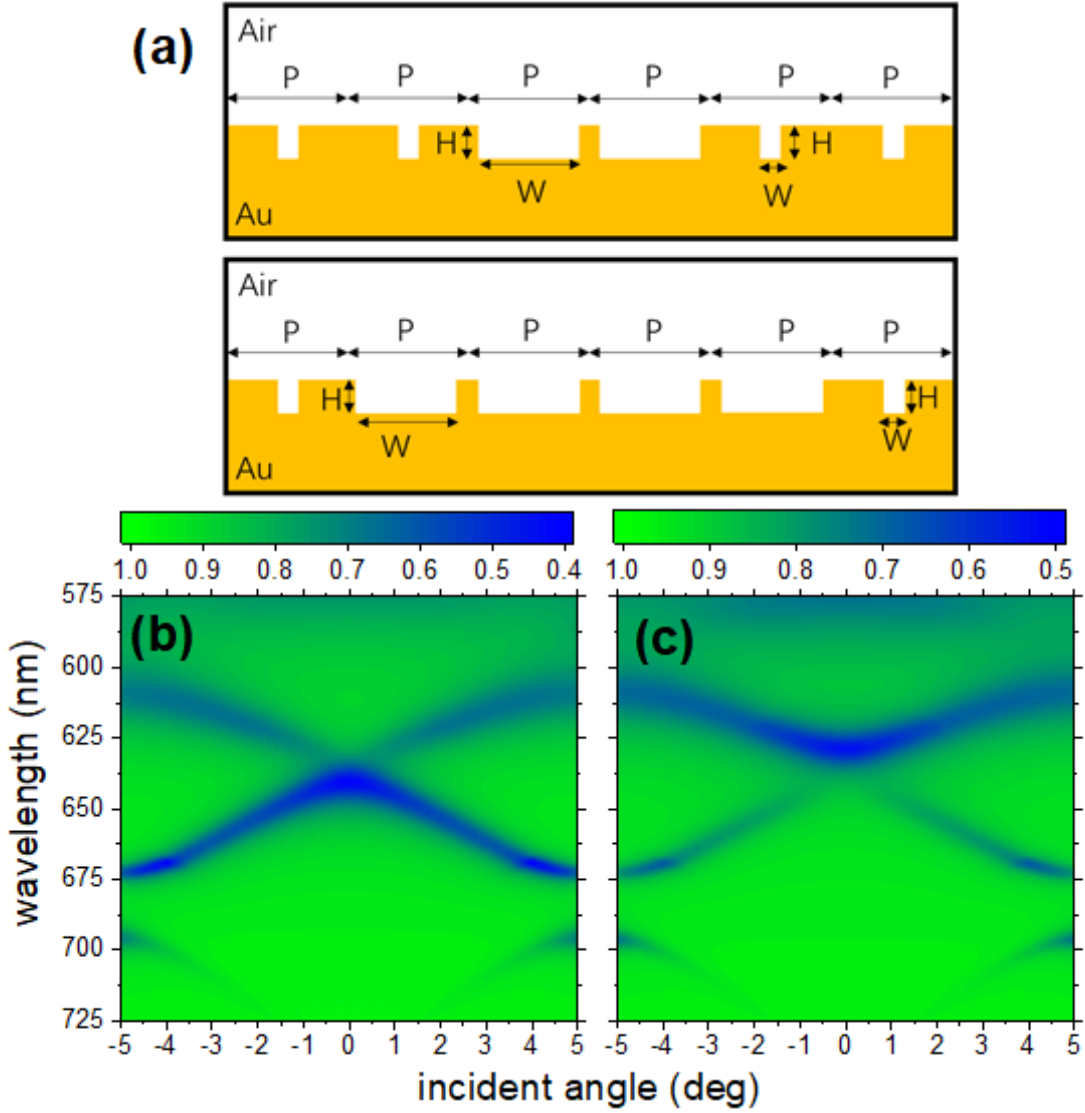

Fig. S6. (a) The FDTD simulation cells of the (top) 2T/2N/2T and (bottom) 1T/4N/1T bipartite superlattices. The angle resolved reflectivity mappings of (b) 2T/2N/2T and (c) 1T/4N/1T superlattices taken along the Γ-X direction.

## G. Simulated near field patterns of the lower interface band of the bipartite superlattices across the first Brillouin zone

By using the dipole source excitation method, the complex near-field patterns of the lower interface band of 2T/2N/2T and 1T/4N/1T bipartite superlattices are simulated. The real and imaginary parts of the surface normal field components, Re($E_z$) and Im($E_z$), taken at 20 nm above the surface across the Brillouin zone from k = -π/P to π/P μm$^{-1}$ are shown in Fig. S7. They will then be used for determining the Zak phase by the Wilson loop method.

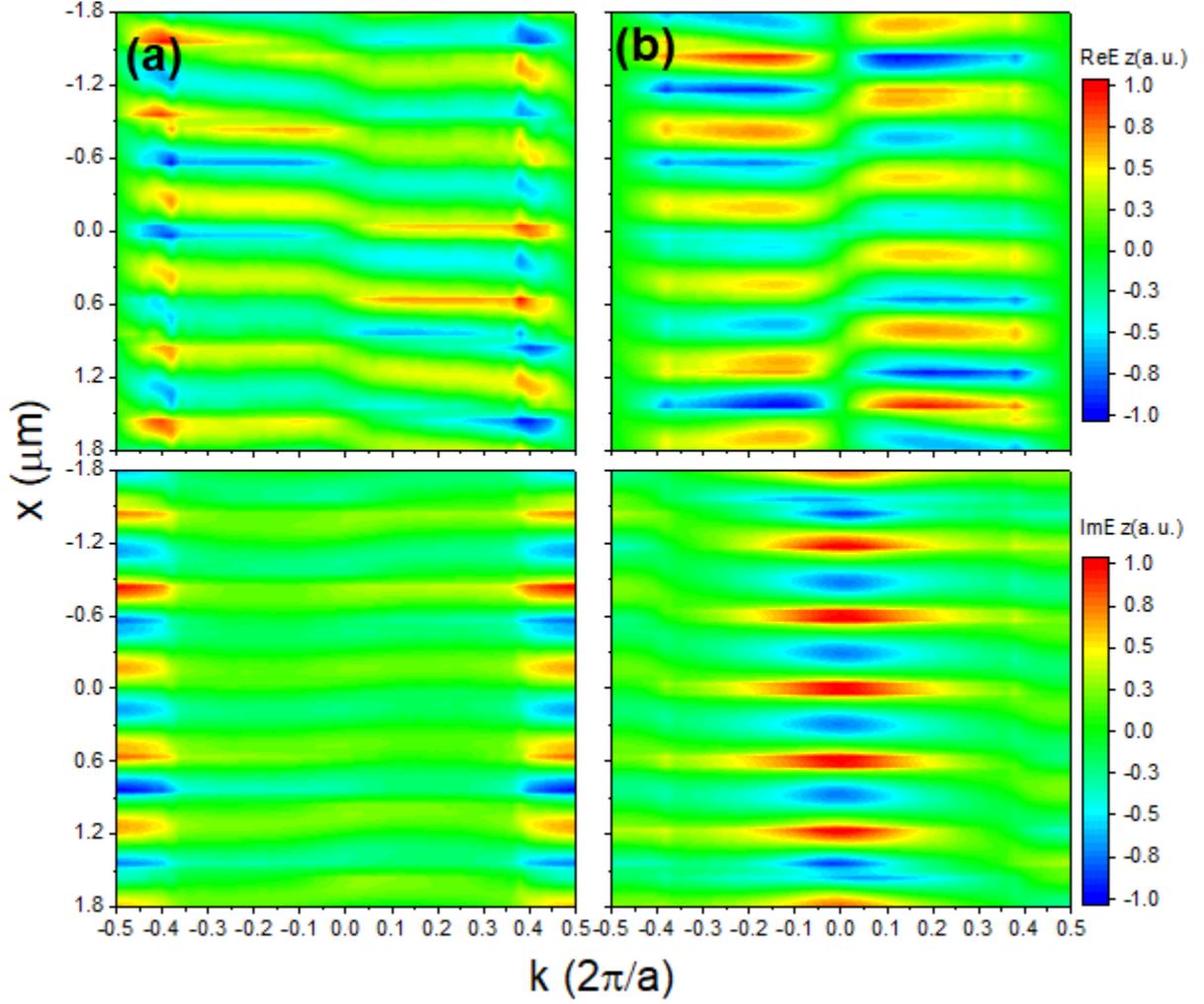

Fig. S7. The (top) real and (bottom) imaginary parts of the surface normal field components of (a) 2T/2N/2T and (b) 1T/4N/1T superlattices across the Brillouin zone.

## H.  Angular profiles of the superlattice interface state

The experimental angular profile of the interface state in the superlattice heterostructure taken at $\lambda = 634$ nm is shown in Fig. S8(a). It is fitted with a Lorentzian function to determine the FWHM to be $\Delta\theta = 1.64$ deg. The simulated angular profile of the interface state in the superlattice heterostructure taken at $\lambda = 635$ nm is shown in Fig. S8(b). It is fitted with a Lorentzian function to determine the FWHM to be $\Delta\theta = 2.29$ deg.

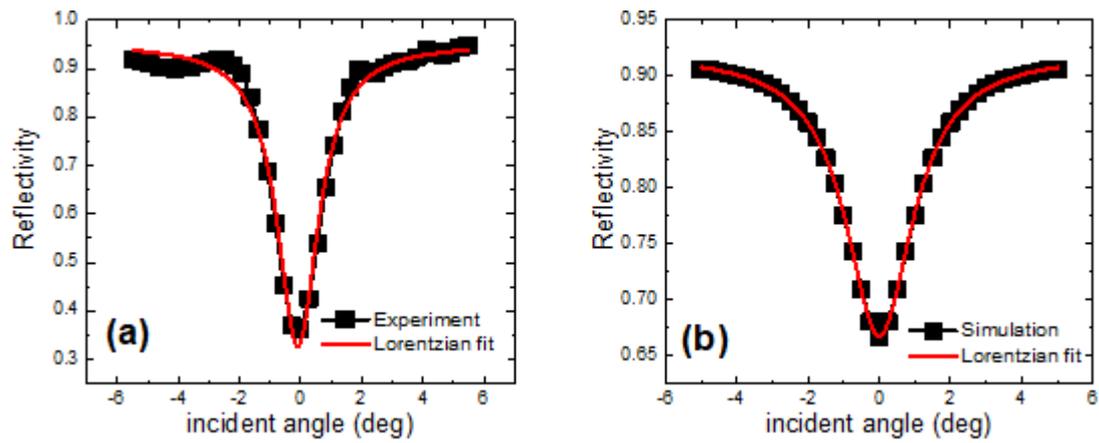

Fig. S8. The (a) experimental and (b) simulated angular profiles of the superlattice interface state.